\documentclass[reprint]{revtex4-1}
\usepackage{graphicx}
\usepackage{dcolumn}
\usepackage{bm}
\usepackage{amsmath}
\usepackage{amssymb}
\usepackage{braket}
\usepackage[utf8]{inputenc}

\begin{document}

\preprint{}

\title{Fault tolerance estimation in digital circuits with visualised generative networks}

\author{Sascha Biel}
\affiliation{IU Internationale Hochschule, Juri-Gagarin-Ring 152, D-99084 Erfurt, Germany}

\author{Carl Alexander Gaede}
\affiliation{IU Internationale Hochschule, Juri-Gagarin-Ring 152, D-99084 Erfurt, Germany}

\author{Amiel Glaser}
\affiliation{IU Internationale Hochschule, Juri-Gagarin-Ring 152, D-99084 Erfurt, Germany}

\author{Jan Wolter}
\affiliation{IU Internationale Hochschule, Juri-Gagarin-Ring 152, D-99084 Erfurt, Germany}

\author{Dr.~Alexej~Schelle}
\affiliation{Constructor University, Bremen gGmbH, Campus Ring 1, 28759 Bremen, Germany}
\affiliation{IU Internationale Hochschule, Juri-Gagarin-Ring 152, D-99084 Erfurt, Germany}

\date{\today}

\begin{abstract}

We propose a new numerical method to estimate the fault tolerance of failure modes in digital circuit structures with a generative network sampling technique.
From a random input of generated bitwise configurations of ideally digitalised analog currents in the digital circuit design with classical logical gates, expected output currents are compared to the realistic signals of a numerical experiment at the discriminator part of the Generative Adversarial Network (GAN) to calculate the deviation from ideal digital electronic signals, including various error modes, such as missing or interchanged logical devices. 
From the present analysis of a representation of the GAN in terms of complex variables, it is possible to evaluate the robustness in electronic designs by differentiating the impact of failure modes associated with different classical logical elements in the circuit. 

\begin{description}
\item[Purpose] Preprint for publication.
\end{description}
\end{abstract}

\maketitle

\section{Introduction}

Generative deep learning networks are used in many areas of modern information technology, such as computer-based art, public media, and synthetic data generation, as well as for creating large language models, source code designs, and game development structures.
The very basics of generative network modeling are to build sampling strategies using two competing networks that interact constructively to intelligently generate new outputs from randomly generated input data \cite{Bhat2025, Shao2023, Ambrogioni2024}.
In a Generative Adversarial Network (GAN) setup \cite{Goodfellow2014, Goodfellow2016}, the generator samples synthetic data recursively using a specific learning strategy and compares it against the expected outcomes at the discriminator of the generative network.
Minimal components of the generative network are built from a single initial input layer and a minimal output reference layer that compares the network's results with expected numerical or experimental results, with a modulator between them.
Intelligent classification is achieved by identifying the correct label classes of structured and unstructured data using supervised learning, by solving coupled multidimensional non-linear equations.

As we have shown in Ref. \cite{Schelle2025} using a reversible circuit scheme, signal processing of high-dimensional binary keys, in particular generative network sampling, allows secure textmail encryption with Generative Adversarial Networks. 
In the reversible GAN design, randomly generating N-dimensional binary keys at the generator enables the calculation of integer numbers that are used for asymmetric encryption and decryption of the cryptographic text strings that are transfered from the sender to the receiver within a perfect encryption scheme, i.e., a direct mapping of either one key value per letter, and the encryption and decryption process performed by sending and storing two different key values that are related by a bijective mathematical mapping defined by a reversible classical circuit.
As a counterpart to that analytic processing procedure for reversible circuit designs, pulsed binary combs shall be applied to non-reversible circuit structures in this study to cache the functionality of logical circuit devices, by comparing the generated output signals in a (numerical) experiment with the ideal signals expected for the particular circuit design \cite{SpringerGAN2024, IEEEGAN2023}.
Distributions of integer numbers in complex space thus allow us to build a new sensor technology from numerical sampling of digital configuration deviations in terms of complex numbers that enables a visualisation of failure mechanisms also in generally non-reversible digital circuits.

\begin{figure}[t]
    \centering
    \includegraphics[width=0.45\textwidth]{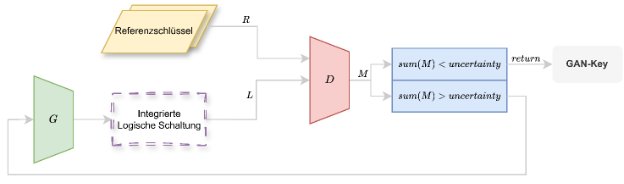} 
    \caption{(Color online) Figures show the GAN setup of the present model for the parsing of erroneous configurations in the connected circuit design. Binary signals are generated randomly at the generator, further modulated in the circuit design, and finally compared at the differentiator to the output signal expected for an ideally non-erroneous circuit configuration at a pre-defined uncertainty level. For deviations exceeding the uncertainty level, the sampling is iterated recursively until the minimum threshold is met. The calculation of the final GAN key at each iteration of the algorithmic sampling is optimized by an 
intelligent modulator that recognizes and stores the bitwise configurations that have provided the expected output signals to accelerate the algorithmic modeling (compare Fig. \ref{figure_2}).}
     \label{figure_1}
\end{figure}

In this paper, we introduce an automated calculation method for modeling fault-tolerance estimation in digital circuit designs using elementary generative artificial intelligence networks, based on the Hopfield energy operator \cite{Ramsauer2020}.  
Analyzing visualisations of the complex-number distribution of model deviations relative to expected (experimental) results yields estimates of the fault tolerance of the circuit designs as a function of different failure modes, such as missing or interchanged logical devices.
We built this modeling technique and the presented analysis as a foundation for the sampling of input data for developing new deep learning artificial intelligence models that can classify the error mode within a digital circuit as a function of the input distribution of complex variables after training the model with training data obtained from the presented GAN sampling method for digital circuits.    
The presented algorithm, in particular, builds a primary idea for a new neuronal network theory that possibly connects elements of neuronal deep learning from the Hopfield theory to the quantum statistical theory of coherently interacting neuronal networks.   

\begin{table}[t]
\centering
\begin{tabular}{c l l}
\hline
Nr. & Bezeichnung & Ausdruck \\ 
\hline
(1)  & AND-NOT & $\overline{A \cdot B}$ \\ 
(2)  & AND-OR & $AB + CD$ \\ 
(3)  & AND-NAND & $\overline{AB \cdot CD} = \overline{AB} + \overline{CD}$ \\ 
(4)  & AND-NOR & $\overline{AB + CD}$ \\ 
(5)  & AND-XOR  & $AB \oplus CD$ \\ 
\hline
(6)  & NOT-AND & $\overline{A} \cdot \overline{B}$ \\ 
(7)  & OR-AND & $(A + B)(C + D)$ \\ 
(8)  & NAND-AND & $(\overline{AB})(\overline{CD})$ \\ 
(9)  & NOR-AND & $(\overline{A + B})(\overline{C + D})$ \\ 
(10) & XOR-AND & $(A \oplus B)(C \oplus D)$ \\ 
\hline
\end{tabular}
\caption{The table shows a comparison of the Boolean logical expressions for reversed logical operations and indicates that an erroneous interchange of electronic devices leads to deviations from the expected binary electronic signals. Numerical sampling of binary configurations, e.g., shows that the logical combinations AND-OR, AND-NAND, and AND-XOR have the most significant impact on failure modes, with up to seventy per cent relative deviation, as compared to the combinations AND-NOT and AND-NOR, which provide maximal relative deviations of up to fifty per cent.}
\end{table}

\begin{figure}[t]
    \centering
   \includegraphics[width=0.35\textwidth]{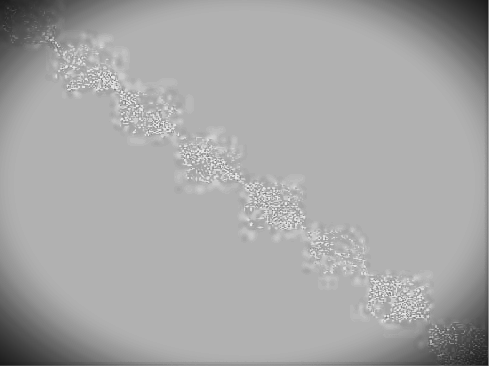} 
    \caption{(Color online) The figure illustrates a representative probabilistic distribution of integer numbers (for a pairwise NOT gate connection between the input bits in a generative and coherent neuronal network) that indicates the deviations of the numerically simulated digital electronic signals from the expected signals for an ideally unperturbed circuit design in a black-and-white depiction of an excerpt from a figure or a diagram. Signatures of the deviations are uniquely visualized by the fluctuations around the expected linear scaling of the difference measurement function $m(f(G), R)$. The fault-tolerance of a circuit design can be estimated by quantifying the uncertainty in transitioning from linear scaling behavior to chaotically distributed probability structures, as a function of the uncertainty level at the transition. }
     \label{figure_2}
\end{figure}

\section{Theory}

We formally map the hybrid GAN topology, consisting of a generator and a discriminator, onto the energy operator of an elementary Hopfield network with one neuronal layer and one reference layer that interact in a symmetric configuration.
To study the implications of the present numerical model in terms of a spectral analysis provided by the Hopfield representation of the Generative Adversarial Network, we consider variables $\xi_i$ and $\xi_j$ (zero for deviation of the simulated bit to the expected bit of a realization of the GAN and one for equal outcomes) for the representations and correlations of the bitwise deviations between expected and instantaneous (experimental) binary signals in the circuit design for a given configuration $\mu$ of the GAN. 
This way, we find that the following eigenvalue equation of the Hopfield energy operator can represent the GAN

\begin{equation}
 \hat{H}  \ket{\xi^\mu_i} \otimes \ket{\xi^\mu_j} = E^\mu_{ij} \ket{\xi^\mu_i} \otimes \ket{\xi^\mu_j}
\end{equation}\\
with eigenenergies

\begin{equation}
 E^\mu_{ij}  = -\frac{1} {2N} \sum^{P}_{\nu = 1}(\xi^\mu_i\xi^\nu_j)^2 \ ,
\end{equation}
where the labels $i$ and $j$ represent neuronal components (bits) and $\mu$ a certain relative configuration of the Generative Adverserial Network \cite{hochreiter2023hopfield, kim2024gan_fault, zhang2025complex}.
Random sampling over all possible configurations $P$ in the limit where $P\rightarrow\infty$ at a given uncertainty level of the deviations (maximally allowed deviations between the input and the output bits) leads to the approximate equation

\begin{equation}
\hat{H} \ket{\xi_i^\mu} \otimes \ket{\xi_j^\mu} = E^\mu \ket{\xi_i^\mu} \otimes \ket{\xi_j^\mu}
\label{eigenvalue}
\end{equation}\\
with eigenenergies that are approximately independent of the neuronal components.
Equation \ref{eigenvalue} represents manifolds of fixed relative uncertainty and projects the energy operator to an observable of the form $\hat{H} =  \sum^{ \infty}_{\mu = 1} E_{\mu} \ket{\xi^\mu} \bra{\xi^\mu}$ with state vectors $\ket{\xi^\mu}$ that are linear combinations of the states $\ket{\xi_i^\mu}\otimes\ket{\xi_j^\mu}$ \cite{wang2026gan_reliability, fernandez2024hopfield_energy}.
Projecting Eq. (\ref{eigenvalue}) onto the basis $\ket{0_i^\mu} \otimes \ket{0_j^\mu}$, $ \ket{0_i^\mu} \otimes \ket{1_j^\mu}$, $ \ket{1_i^\mu} \otimes \ket{0_j^\mu}$ and $ \ket{1_i^\mu} \otimes \ket{1_j^\mu}$, leads to the classical limit within an N-dimensional vector representation, which we consider in the following, in particular, neglecting effects of finite temperature.

\begin{figure}[t]
    \centering
    \includegraphics[width=0.40\textwidth]{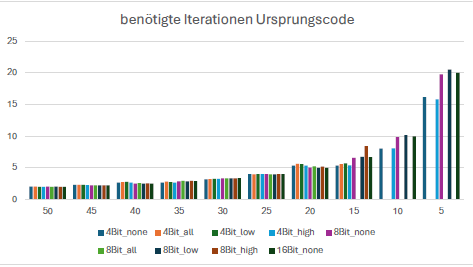} 
    \caption{(Color online) Figure shows the exponential scaling behavior of the number of required iteration steps as a function of the uncertainty level. This scaling behavior is basically independent of the distribution of elementary classical logical gates (AND, OR, NOT, NAND, NOR, XOR, XNOR) that were distributed randomly over the bitwise strings of the simulation model. The number of required iteration cycles, hence, mainly increases with the number of algorithmic steps required for the convergence of the binary vectors to an expected reference signal in a black-and-white depiction of an excerpt from a figure or a diagram.}
     \label{figure_3}
\end{figure}

In this classicial limit, configurations of neuronal GANs can be represented as a statistical ensemble of compound complex numbers with a real part built from a linear combination of binary weighting coefficients at the generator and an imaginary part that represents the differentiator with weighting factors, both in terms of the basis elements $2^j$ for connecting binary numbers $\mathbb{Z}_2$ to integer values $\mathbb{N}$. 
The so-obtained distribution of complex numbers highlights configurations of the GAN as a complex-valued probability distribution in the complex plane.
Applying this scenario to classical circuits that are modeled by ideal binary configurations of analog current modes with connected standard logical gates (see Fig. \ref{figure_3}), an N-dimensional binary vector obtained from one realization of random sampling is mapped with a not necessarily bijective function from the input configuration to an equal-dimensional output vector.

From the GAN setup of Fig.  \ref{figure_1}, and from the Hopfield energy operator for subspaces of constant signal uncertainty, we assert that visualisations of fault effects in the connected digital circuit can be classified within elementary mathematical substructures 
by the mathematical relation

\begin{equation}
m\left(f(G), R\right) = \sum^{N}_{j=1} f_m(a_j) 2^j + i \sum^{N}_{j=1} b_j 2^j ,
\label{deviations}
\end{equation}\\
that defines the deviations of the signal results from expected outcomes.
In Eq. (\ref{deviations}), $a_j, b_j\in\mathbb{Z}_2$ are coefficients that describe relations between the generator and the differentiator of the generative network (the difference of these coefficients builds the variables $\xi_i$). 
The function $f:\mathbb{Z}_2 \rightarrow \mathbb{Z}_2$ in particular accounts for the bitwise modulations $f_m(a_j)$ of the $j$th component in an N-dimensional 
binary signal in the digital circuit design that is considered to be generally not reversible in the present study.
Recursive calculations of $m\left(f(G), R\right)$ at a predefined level of relative uncertainty between the input and the expected integer representations of the input and the expected binary signals, lead to the probabilistic distribution of deviations between effectively generated signals at the generator and the differentiator, see Figs. \ref{figure_2}.  

\begin{figure}[b]
    \centering
    \includegraphics[width=0.15\textwidth]{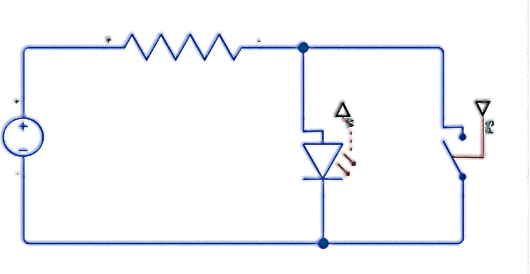} 
    \includegraphics[width=0.15\textwidth]{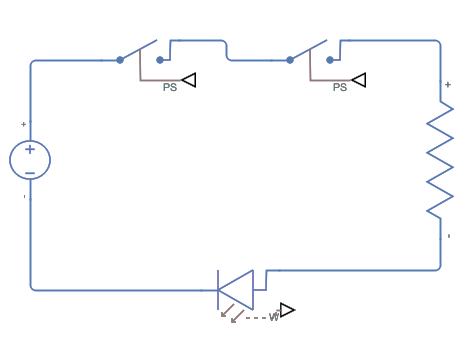} 
    \newline
    \includegraphics[width=0.15\textwidth]{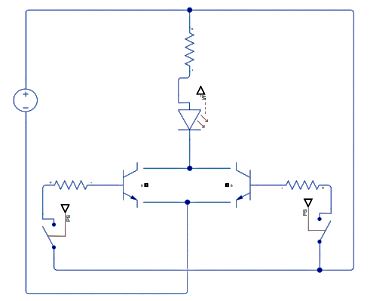} 
    \includegraphics[width=0.15\textwidth]{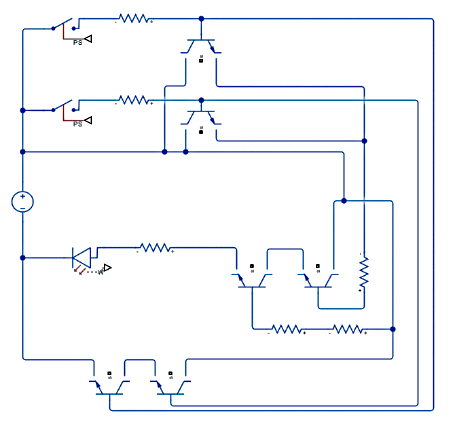} 
    \newline
    \includegraphics[width=0.15\textwidth]{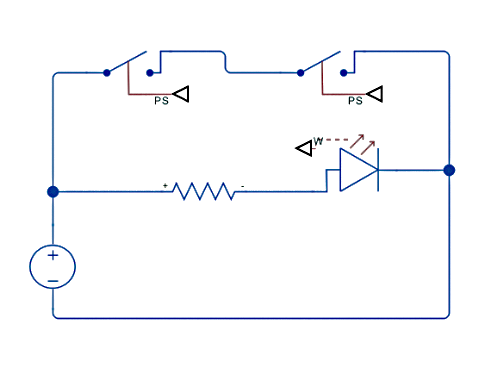} 
    \includegraphics[width=0.15\textwidth]{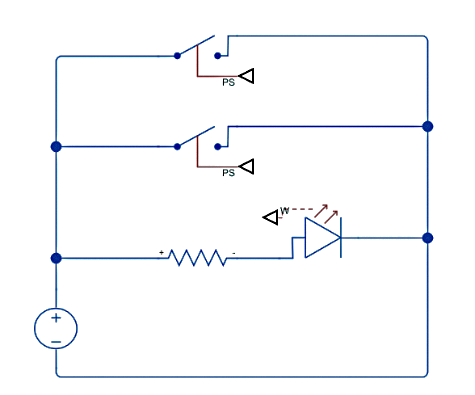} 
    \newline
    \includegraphics[width=0.15\textwidth]{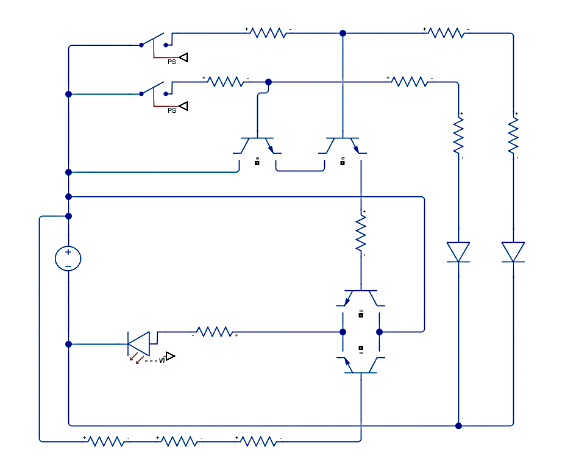} 
    \caption{(Color online) There are different ways to construct digital logical gates from analog electronic devices. 
In this figure, we depict one possible strategy of mapped analog circuits onto digital logical devices using predefined levels of analog voltages and currents for swichting, such as to model classical digitial NOT (upper left) , AND (upper right), OR (middle left), XOR (middle right), NAND (lower left),  NOR (lower right), and  XNOR (last) gates.}
     \label{figure_4}
\end{figure}

\begin{figure}[t]
    \centering
    \fbox{\includegraphics[width=0.20\textwidth]{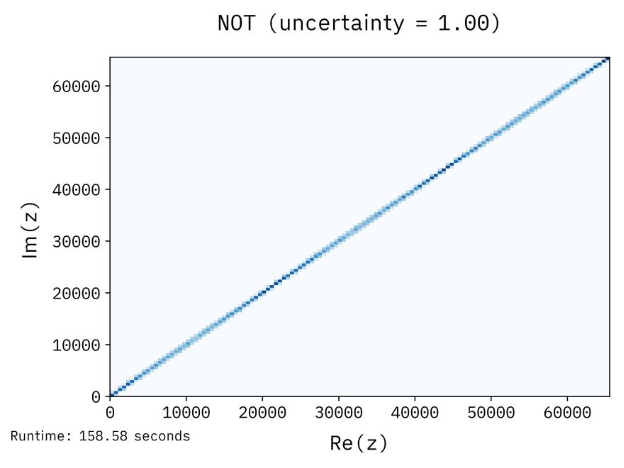}}\hspace{0.01\textwidth}
    \fbox{\includegraphics[width=0.20\textwidth]{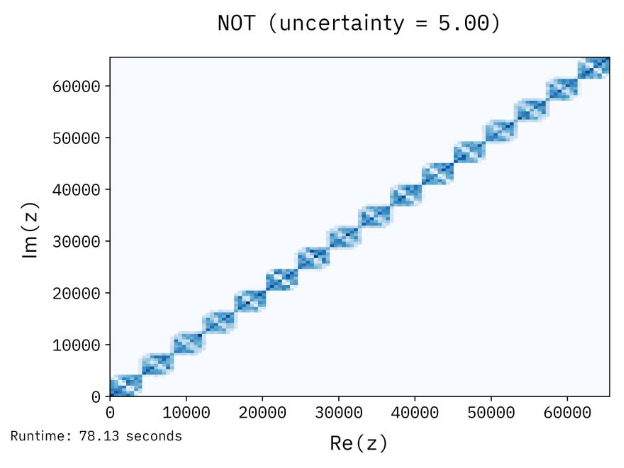}}
    \newline
    \fbox{\includegraphics[width=0.20\textwidth]{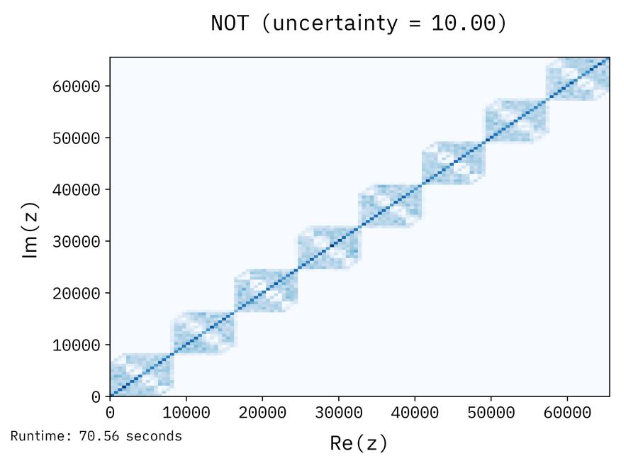}}\hspace{0.01\textwidth}
    \fbox{\includegraphics[width=0.20\textwidth]{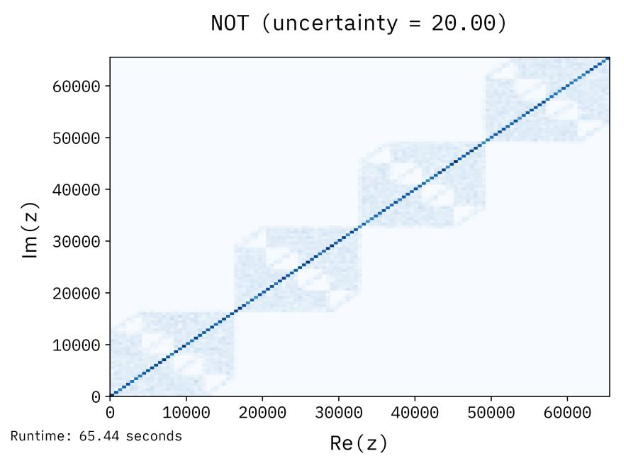}}
    \caption{(Color online) Figure illustrates the transition from linear scaling behavior of pairwise connected logical NOT gates in a 16-bit neuronal network with a single input layer (GAN). Transition from linear scaling to chaotic probability distribution occurs at about figure 3, which corresponds to a ten percent deviation of the expected from the required ideal output signal. We hence conclude that standard logical NOT gates show a fault tolerance of about ten percent until a complete failure of the overall circuit design. In contrast to NOT logical gates, OR and XOR logical gates turn out to be more robust against errors and less convergent in a GAN digital circuit design.}
    \label{figure_5}
\end{figure}

In the present study, we have considered three main error modes in digital circuit designs: missing devices or perturbations in the input signal (that define equivalent error modes in a single-layer neuron model as considered here), 
and reversed-polarity components in the circuit for standard classical logical \cite{chen2026gan_sampling}. 
With our theoretical approach, we also draw attention to the robustness of certain circuit designs as a possible applicational scope of the algorithm \cite{ li2025robustness}.
Fault-tolerance is estimated from the uncertainty level at which a transition occurs from linear to non-linear scaling behaviour of the configuration deviations in the representation of complex numbers.     
In the GAN setup, erroneous configurations in the circuit design are tested with binary signals that are generated probabilistically at the generator, and further modulated in the circuit design until the final comparison at the differentiator. 
Deviations exceeding the defined uncertainty level are iterated until the minimum threshold is met. 
The scaling of the number of required iterations as a function of the defined relative uncertainty is shown in Fig. \ref{figure_3}.

\begin{figure}[t]
    \centering
    \fbox{\includegraphics[width=0.45\textwidth]{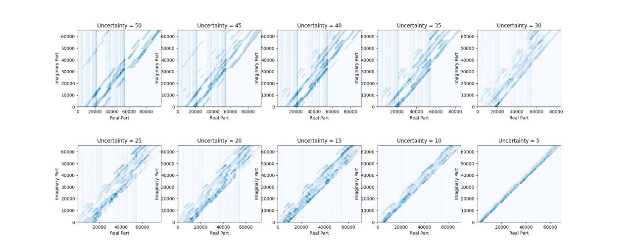}}\hspace{0.01\textwidth}
    \fbox{\includegraphics[width=0.45\textwidth]{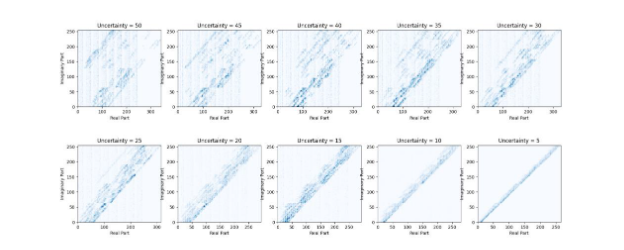}}\hspace{0.01\textwidth}
    \caption{(Color online) The transition from linear scaling to deviations from standard linearity is shown in the figure for a combination of AND and NOT logical gates for an 8-bit (lower figure) and a 16-bit neuronal network layer (upper figure). For the 8-bit layer, the distribution shows larger deviations compared to the 16-bit neuronal network layer. A transition to clearly nonlinear behavior occurs at about fifteen to twenty per cent deviation. We have observed similar scaling behavior for other combinations of standard logical gates.} 
    \label{figure_6}
\end{figure}

\section{Results}

Within our Generative Adversarial sampling approach, we have simulated neuronal networks with 4, 8, 12, and 16-bit configurations \cite{ramirez2025bitwidth}.  
Calculation time required for convergent results ranges from about a few seconds of computation time for 4-bit configurations to about several minutes and hours for 16-bit layers  \cite{schmidt2024scaling}.
We have tested standard logical AND, OR, NOT, NAND, NOR, XOR, and XNOR gates, with approximated analog circuits that are modeled by digital circuit designs.
As shown in Fig. \ref{figure_4}, specific designs of analog circuits can be mapped and approximated with digital circuit designs, assuming offsets for the switching level from bitwise 0 and 1 configurations, 
which is an analog current below or above the defined threshold. 
Standard logical NOT gates are equivalent to analog circuits that show a parallel switch to the considered paths; if the switch is turned on, no current flows towards the relevant path (represented by a light-emitting diode), whereas 
for configurations with disconnected switches, the analog NOT configuration (upper left circuit in Fig. \ref{figure_4}), shows a current above a certain voltage and current level, above which the diode is emitting incoherent light. 
The upper right analog circuit design represents a logical AND gate that leads to a logical value of 1 if both switches are turned on, and to the logical value of 0 if both or only either or both switches are turned on. 
A logical OR gate can be represented by the analog circuit design of the middle left panel, where the switching of either one or both of the switches leads to currents to modulate the transistors of either one or both of the two possible paths in the circuit.
Digital XOR gates can be presented by the circuit design in the middle right of Fig. \ref{figure_4}, where an XOR gate is shown. 
The diode only emits light when exactly one of the two switches is closed. 
If neither switch is closed, the bottom-right transistor does not turn on, and the current path through the diode remains open. 
If both switches are closed, both transistors at the bottom left turn on, and the current flows through them instead of turning on the second transistor at the bottom right, because a higher resistance (symbolically represented here by two resistors) is placed before that transistor. 
If only one of the two switches is closed, the first transistor at the bottom right is turned on by the signal from that switch, and the second transistor remains on as long as neither of the transistors at the bottom left is turned on.
A logical NAND gate is obtained from the analog AND configuration with a diode that is placed in parallel instead of in series to the light-emitting diode.
We could also verify that logical NOR gates can be modeled with two switches that are connected in parallel to the light-emitting device.
Finally, we have built an analog counterpart for XNOR logical gates that are equivalent to connecting electronic devices, as shown in the lowest sketch in Fig. \ref{figure_4}.
The light-emitting diode is only actively emitting energy if either of the switches or none of the switches is closed.

Replacing analog circuits with approximated digital circuit designs using the components shown in Fig. \ref{figure_4} to map the analog to the digital design enables the estimation of fault tolerance by finding the uncertainty levels at which a visible transition occurs from linear scaling to chaotic structures of the probability distribution for the deviation function $m(f(G), R)$.
As represented in Fig. \ref{figure_5}, a transition from linear to clearly non-linear scaling for a 16-bit neuronal network with only one input layer and pairwise NOT connections occurs at about ten to twenty percent fault tolerance (deviation of the simulated output signal from the expected output signal, where the maximum predefined uncertainty of the relative configurations is reached).
We have performed the same analysis for classical AND, OR, NAND, NOR, XOR, and XNOR gates, which have indicated that AND and NAND gates require more iteration steps for convergence than OR, NOR, XOR, and XNOR gates for predefined uncertainty levels.
In the above setup, we have primarily modeled the scaling behavior in terms of deviations between the current and expected signals, which resemble the practical case of missing devices or erroneous input signals in the experimental circuit design.  

In Fig. \ref{figure_5}, deviations from linearity appear in the combination of AND and NOT as a consequence of erroneously interchanged devices. 
We have tested all possible combinations of pairwise connections and interchanges of logical gates and find that combinations of AND, OR, and XOR logical gates show more robust scaling behavior than 
AND, OR, and XOR logical gates that are connected to NOT logical gates.
Transition of clearly linear to non-linear scaling occurs at a ten to fifteen percent uncertainty level of the binary signals.
It is also visible that the 8-bit configurations of the neuronal layer show larger deviations at the same uncertainty level as compared to a 16-bit neuronal network layer, starting at an uncertainty level of about thirty percent.

\begin{figure}[b]
    \centering
    \fbox{\includegraphics[width=0.20\textwidth]{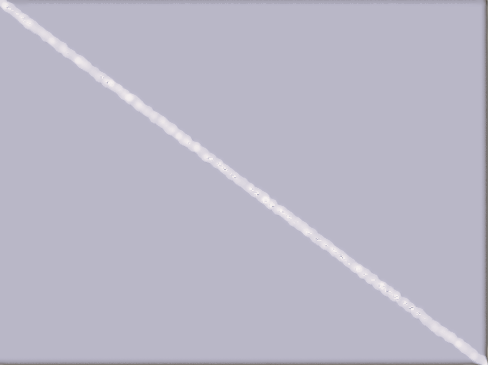}}\hspace{0.01\textwidth}
    \fbox{\includegraphics[width=0.20\textwidth]{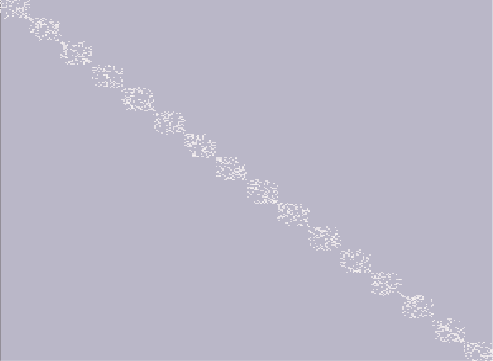}}\hspace{0.01\textwidth}
    \fbox{\includegraphics[width=0.20\textwidth]{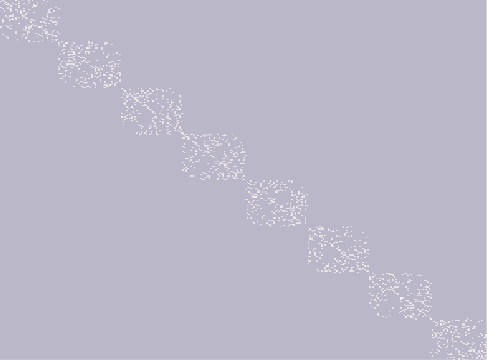}}\hspace{0.01\textwidth}
    \fbox{\includegraphics[width=0.20\textwidth]{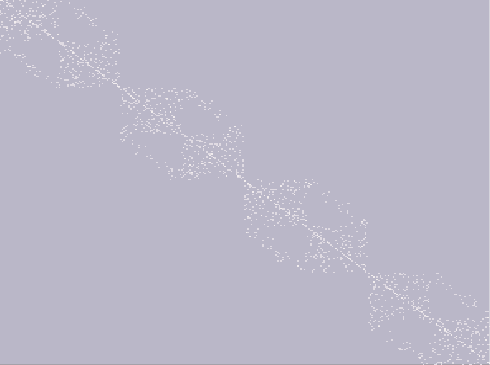}}\hspace{0.01\textwidth}
    \caption{(Color online) Extracting unstructured picture formats as shown above from the probabilistic distribution of the deviation function leads to data sources required for building new deep learning 
artificial intelligence models that are trained in terms of the probability distribution structure and the classifying labels of the failure modes in a supervised learning mode to predict the failure mode of a 
experimentally tested circuit design.} 
    \label{figure_7}
\end{figure}

\section{Discussion}

As we understand from the Hopfield ansatz, finding configurations with minimal and constant deviations between expected and erroneous binary signal configurations corresponds to stable states with minimal total energy of the neuronal network.
Collecting all configurations of definite uncertainty provides a complete system with eigenstates and eigenenergies that represent the entire space of possible configurations.
Subspaces of different uncertainty levels or different logical modes and circuit designs are not orthogonal in general, but so are the eigenstates of different configurations at a given definite uncertainty level.

From the estimation of the transition levels at which linear scaling of the model deviations to expected configurations of binary circuit designs leads to definite values of fault tolerance estimate for circuit designs, that is, the level at which one expects the electronic devices to entirely fail in application, whereas well below that level, small errors may be tolerated in practice.
Comparing different elementary circuit structures with pairwise-connected standard logical gates (AND, OR, NAND, NOR, NOT, XOR, XNOR), we find that probability structures for deviations in model results can be differentiated across circuit designs and logical gates.
Connecting both the visualized deviations and the spectral structure of the model system leads to the possibility of manually analyzing the fault tolerance of circuit designs. 

Regarding the convergence of the GAN sampling strategy, we find that the percentage of all configurations of final N-dimensional binary vectors that can be resembled with modulated input binary structures varies with the type of elementary logical gates, the maximally allowed uncertainty level 
and the maximum number of allowed iteration steps.
Standard classical NOT gates show about a factor of two to three times faster convergence than the classical AND, NAND, NOR, OR, or XOR logical gates, which is visible in a reduced intensity of the probabilistic distribution.
As the complexity of the circuit design increases, convergence is more difficult to obtain, and the uncertainty level must be increased to allow convergent results.

In our current study, we would like to mention the possibility of developing deep learning artificial intelligence models that enable the automated classification of failure modes from the input of probabilistic representations and visualizations of the deviations for binary configurations in future works \cite{fischer2025manual_vs_auto}.
Within our approach, we were able to build the foundation for the estimation of fault tolerance and to generate a model for generative input data to support deep learning models from visualizations of probabilistic model deviations.
As indicated by the representation of neuronal networks within simple models like the Hopfield theory of single-layer neuronal networks, as explained, we assume that automized classifications can not trivially be deduced from a standalone analysis of the GAN results, but require more complex artificial intelligence models that analyse both the real-valued spectrum from the Hopfield theory and the detailed substructure of the GAN probabilistic representation of model deviations in complex space.
For that purpose, coherent field configurations of hybrid atomic states \cite{schelle2023comb}, taking into account also the effects of finite temperature and the modulations from the distribution of chemical potentials associated with the Hopfield energetic spectrum in the limit of quantum mechanical scales, should be considered.
  
\section{Conclusion}

In the present study, we have illustrated a new approach to building digital circuit designs from analog electronic structures to estimate the fault tolerance in elementary classical circuit designs by defining the transition of linear scaling to non-linear deviations of empirical model results to theoretical scaling.
The scaling behavior changes from linear to chaotic probability structures in complex space at around ten to twenty percent relative deviations for missing devices and erroneous input signals, and about twenty to thirty percent for interchanged logical devices. 
The presented simulation algorithm builds the foundation for the application and modeling of more complex circuit designs as a counterpart for the numerical sampling of error modes in generally non-reversible circuit designs. 
With our analysis of a new Hopfield ansatz for GANs, we have proven the completeness of eigenstates in subspaces of constant signal uncertainty to build the foundation for data sampling to train new deep learning artificial intelligence models for accurate error mode classification.

\acknowledgments
Dr. A. Schelle acknowledges the financial support from IU Internationale Hochschule for the continuous and ongoing (freelancer) lecturer position at the university.  
I also thank Hauke Debbeler, Jonas Kinas, Sebastian Neuperger, Aleksa Percic, and Daniel Wolf for their contributions and discussion regarding the content of the presented work on the fault-tolerance estimation in digital circuits using generative network sampling.

\bibliography{references}     


\end{document}